\documentclass[amsmath, floatfix,twocolumn, tightenlines, superscriptaddress, amssymb, aps, prl]{revtex4-2}

\usepackage{graphicx}
\usepackage{dcolumn}
\usepackage{bm}
\usepackage[colorlinks=true, linkcolor=blue, citecolor=blue]{hyperref}
\usepackage[table,usenames,dvipsnames,svgnames]{xcolor}
\usepackage{pifont}
\usepackage{enumitem}
\usepackage[normalem]{ulem}

\usepackage{booktabs}
\usepackage{multirow}

\newcommand\ip[1]{\left\langle#1\right\rangle}
\newcommand\ket[1]{\left|#1\right\rangle}

\newcommand\op[2]{\left|#1\right\rangle\!\left\langle#2\right|}
\newcommand\id{\mathbb{I}}
\DeclareMathOperator\tr{Tr}
\DeclareMathOperator*{\Prob}{Prob}

\newcommand{\error}{\varepsilon}
\newcommand{\jnrisk}{\varepsilon_*}
\newcommand{\failure}{\delta}

\begin{document}

\title{Versatile fidelity estimation with confidence}

\author{Akshay Seshadri}
\affiliation{Department of Physics, University of Colorado Boulder, Boulder, USA}
\author{Martin Ringbauer}
\affiliation{Universit\"{a}t Innsbruck, Institut f\"{u}r Experimentalphysik, Technikerstrasse 25, 6020 Innsbruck, Austria}
\author{Jacob Spainhour}
\affiliation{Department of Applied Mathematics, University of Colorado Boulder, Boulder, USA}
\author{Rainer Blatt}
\affiliation{Universit\"{a}t Innsbruck, Institut f\"{u}r Experimentalphysik, Technikerstrasse 25, 6020 Innsbruck, Austria}
\affiliation{Institut f\"{u}r Quantenoptik und Quanteninformation, \"{O}sterreichische Akademie der  Wissenschaften, Otto-Hittmair-Platz 1, 6020 Innsbruck, Austria}
\affiliation{Alpine Quantum Technologies GmbH, 6020 Innsbruck, Austria}
\author{Thomas Monz}
\affiliation{Universit\"{a}t Innsbruck, Institut f\"{u}r Experimentalphysik, Technikerstrasse 25, 6020 Innsbruck, Austria}
\affiliation{Alpine Quantum Technologies GmbH, 6020 Innsbruck, Austria}
\author{Stephen Becker}
\affiliation{Department of Applied Mathematics, University of Colorado Boulder, Boulder, USA}

\date{\today}

\begin{abstract}
    As quantum devices become more complex and the requirements on these devices become more demanding, it is crucial to be able to verify the performance of such devices in a scalable and reliable fashion. A cornerstone task in this challenge is quantifying how close an experimentally prepared quantum state is to the desired one. Here we present a method to construct an estimator for the quantum state fidelity that is compatible with any measurement protocol. Our method provides a confidence interval on this estimator that is guaranteed to be nearly minimax optimal for the specified measurement protocol. For a well-chosen measurement scheme, our method is competitive in the number of measurement outcomes required for estimation. We demonstrate our method using simulations and experimental data from a trapped-ion quantum computer and compare the results to state-of-the-art techniques. Our method can be easily extended to estimate the expectation value of any observable, such as entanglement witnesses.
\end{abstract}

\maketitle
With quantum devices starting to push the limit of their classical counterparts, it is becoming increasingly challenging to verify and validate that these devices operate as intended. The gold standard for characterizing quantum states and operations is quantum tomography~\cite{Vogel1989, Hradil1997}. However, tomography, even its more efficient incarnations~\cite{Flammia2012, Gross2010, Cramer2010, Riofrio2017, Lanyon2017}, quickly becomes impractical as the system size grows. Moreover, depending on the kind of reconstruction used, providing reliable uncertainty bounds to the fidelity obtained from a reconstructed quantum state has proven very challenging~\cite{Blume-Kohout2012ErrorBars, Suess2017}.

Fortunately, it is often sufficient to ascertain that the quantum state (or process) in the experiment is close to the desired one, rather than fully reconstructing it. A commonly used measure for this ``closeness" is the fidelity between the two states. For certain classes of states, the fidelity can be estimated from only a few Pauli measurements using a technique called direct fidelity estimation (DFE)~\cite{Flammia2011, DaSilva2011}. For a given target state, DFE specifies a measurement scheme that is to be sampled on the experimental state. However, depending on experimental constraints, the prescribed measurement scheme might not always be the one that is easiest to implement or achieves the highest accuracy. Moreover, the question of how to compute tighter confidence regions for such estimates remains.

Here we propose a method that not only provides fidelity estimates for arbitrary measurement schemes but also guarantees a nearly minimax optimal (symmetric) confidence interval for the chosen measurement scheme and target state. This method, which we denote the \textit{minimax} method, is based on results in statistics by Juditsky \& Nemirovski, who describe the estimation of linear functionals under general conditions~\cite{Juditsky2009, goldenshluger2015hypothesis, juditsky2018near}. In the minimax method, any linear estimator can be constructed for a specified measurement scheme and target state, which can subsequently provide estimates with guaranteed confidence regions from experimental data. Furthermore, in some settings, the performance of the estimator can be precisely characterized in terms of \emph{sample complexity}, which is the minimum number of measurement outcomes necessary to find an estimate within a desired error and confidence level for the chosen measurement setting. For example, we show in the companion paper~\cite{PRA} that for a confidence level $1 - \failure$ our method can produce a fidelity estimate within an additive error, or \emph{risk} (defining a symmetric confidence interval) of $\jnrisk$, with as few as $\sim \ln(2/\failure)/(2\jnrisk^2)$ measurement outcomes for any target state. This sample complexity is achieved when measuring in a basis defined by the target state. In contrast, tomography with Pauli measurements can require exponentially more measurement outcomes~\cite{Flammia2011}. While measuring in the aforementioned basis is often impractical, our method achieves a comparable sample complexity for stabilizer states, like those used in quantum error correction, by randomly sampling elements from the stabilizer group. For more general states, the sample complexity is comparable to that of DFE, when using the measurement scheme described in section II.E. of Ref.~\cite{PRA}, and thus remains practical for typical states.

In the following, we consider an experimenter aiming to characterize an unknown quantum state $\sigma \in \mathcal{X}$ using $L$ different measurement settings, where $\mathcal{X}$ denotes the set of density matrices. Each measurement setting is described by a positive operator-valued measure (POVM) $\{E^{(l)}_1, \dotsc, E^{(l)}_{N_l}\}$ such that $\sum_i E^{(l)}_{i} = \id$ for $l = 1, \dotsc, L$. The $l^\text{th}$ measurement is repeated $R_l$ times (i.e., experimental shots) and the set of possible outcomes $\Omega^{(l)} = \{1, \dotsc, N_l\}$ is labeled by the index of the POVM elements. We denote by $p_\sigma^{(l)}$ the distribution of outcomes from the $l^\text{th}$ measurement setting, given by the Born rule as $p_\sigma^{(l)} = (\tr(E^{(l)}_1 \sigma), \dotsc, \tr(E^{(l)}_{N_l} \sigma))$~\footnote{Technically, our approach requires non-zero probabilities for each outcome, which we omit here for brevity and discuss in detail in Ref.~\cite{PRA}.}. However, since the state $\sigma$ is unknown, so are the distributions $p_\sigma^{(l)}$. Hence, following Ref.~\cite{Juditsky2009}, we start with a family of distributions $p_\chi^{(l)}$ over the outcomes $\Omega^{(l)}$, parametrized by density matrices $\chi \in \mathcal{X}$. 

We now seek to construct an estimator $\widehat{F}_\ast$ that takes measurement outcomes for the specified measurement scheme as an input and returns an estimate for the fidelity between the state $\sigma$ and a pure target state $\rho$ with reliable confidence regions. 
Formally, we define for a given confidence level $1 - \failure \in (0, 1)$ and estimator $\widehat{F}$ an $\failure$-risk $\mathcal{R}(\widehat{F}; \failure)$ as the smallest symmetric interval around $\widehat{F}$ such that the probability that the true value is outside that interval is smaller than $\failure$:
\begin{align}
    \mathcal{R}(\widehat{F}; \failure) = \inf \bigg\{&\error\ \big\vert \sup_{\chi \in \mathcal{X}} \nonumber \\ \Prob_{\text{outcomes} \sim p_\chi}\bigg[ 
    &\left|\widehat{F}(\text{outcomes}) - \tr(\rho \chi)\right| > \error \bigg] < \failure\bigg\}.
    \label{eq:risk}
\end{align}
Here, ``$\text{outcomes} \sim p_\chi$'' means that the outcomes for $l^\text{th}$ measurement are distributed according to $p_\chi^{(l)}$ for a given state $\chi$, for all $l = 1, \dotsc, L$. 
Notably, this risk is a property of the measurement strategy, the target state, the chosen estimator, and the desired confidence interval, but \textit{not} of the actual state that was prepared in the lab. Using this definition, we denote the \textit{minimax optimal} $\failure$-risk as $\mathcal{R}_\ast(\failure) = \inf_{\widehat{F}} \mathcal{R}(\widehat{F}; \failure)$, where the infimum is taken over all real-valued functions over the set of outcomes.

In practice, considering all possible functions as candidate estimators makes the problem computationally unmanageable. To address this challenge we follow Ref.~\cite{Juditsky2009} and consider a restricted set $\mathcal{F}$ consisting of estimators $\phi$ of the form $\phi = \sum_{l = 1}^L \phi^{(l)}$, where $\phi^{(l)} \in \mathcal{F}^{(l)}$ takes an outcome of $l^\text{th}$ POVM as input and returns a number. 
Here, $\mathcal{F}^{(l)}$ is a finite-dimensional vector space consisting of all estimators for the $l$th POVM. The risk achieved when we restrict our estimators to $\mathcal{F}$ is called ``affine'' risk. 
As shown in Ref.~\cite{Juditsky2009}, the restriction to $\mathcal{F}$ is unproblematic, as one can still construct an estimator $\widehat{F}_* \in \mathcal{F}$ which achieves a risk $\jnrisk$ that is almost minimax optimal for $\failure \in (0, 1/4)$ in the sense that:
\begin{align}
    \jnrisk &\leq \vartheta(\failure) \mathcal{R}_*(\failure) \label{eq:almostoptimal}\\
    \vartheta(\failure) &= 2 + \frac{\ln(64)}{\ln(0.25/\failure)} \label{eq:theta}
\end{align}
For confidence levels greater than $90\%$, $\vartheta(\failure)$ is of the order of $1$, and decreases as the confidence level increases. This estimator $\widehat{F}_*$ is constructed as follows:
\begin{enumerate}[leftmargin=0.2cm]
    \item Find the saddle-point value of the concave-convex function
    \begin{align}
        \Phi(\chi_1, \chi_2;&\ \phi, \alpha) = \tr(\rho \chi_1) - \tr(\rho \chi_2) + 2\alpha \ln(2/\failure) \nonumber \\
                                             &+ \alpha \sum_{l = 1}^L R_l \Bigg[\ln\left(\sum_{k = 1}^{N_l} 
                                                           \exp\left(-\phi^{(l)}_k /\alpha\right) \tr(E^{(l)}_k \chi_1)\right) \nonumber \\
                                             &\hspace{1.5cm} + \ln\left(\sum_{k = 1}^{N_l} \exp\left(\phi^{(l)}_k /\alpha\right) 
                                                                \tr(E^{(l)}_k \chi_2)\right)\Bigg]
    \end{align}
    to a given precision, by maximizing over the density matrices $\chi_1, \chi_2 \in \mathcal{X}$ and minimizing over the candidate estimator $\phi \in \mathcal{F}$ and the positive number $\alpha > 0$, see Ref.~\cite{PRA} for details. Since each $\Omega^{(l)}$ is a finite set, each $\phi^{(l)} \in \mathcal{F}^{(l)}$ can be represented as an $N_l$-dimensional real vector with its $k^\text{th}$ component denoted by $\phi^{(l)}_k$.

    \item Denote the saddle-point value of $\Phi$ by $2\jnrisk$, i.e., $2\jnrisk = \inf_{\alpha > 0, \phi \in \mathcal{F}} \max_{\chi_1, \chi_2 \in \mathcal{X}} \Phi(\chi_1, \chi_2; \phi, \alpha)$. Suppose that the saddle-point value is attained at $\chi_1^*, \chi_2^* \in \mathcal{X}$, $\phi_* \in \mathcal{F}$ and $\alpha_* > 0$ to within the given precision. Recall that by definition of $\mathcal{F}$, we have $\phi_* = \sum_{l = 1}^L \phi^{(l)}_*$ with $\phi^{(l)}_* \in \mathcal{F}^{(l)}$.

    \item Let $\{o^{(l)}_1, \dotsc, o^{(l)}_{R_l}\}$ denote $R_l$ independent and identically distributed outcomes obtained for the $l$th measurement setting. Then the estimator $\widehat{F}_* \in \mathcal{F}$ is given as
        \begin{align}
            &\widehat{F}_*(\text{outcomes}) = \sum_{l = 1}^L \sum_{k = 1}^{R_l} \phi^{(l)}_*(o^{(l)}_k) + c
            \label{eq:estimator}
        \end{align}
        with the constant $c = \frac{1}{2} \left(\tr(\rho \chi_1^*) + \tr(\rho \chi_2^*)\right)$.
        The risk associated with this estimator satisfies $\mathcal{R}(\widehat{F}_*; \failure) \leq \jnrisk$, so the final output will be $\widehat{F}_*(\text{outcomes}) \pm \jnrisk$.
\end{enumerate}
An intuitive way to understand the estimator of Eq.~\eqref{eq:estimator} is as a simple weighting of the relative frequencies observed in the experiments (see Ref.~\cite{PRA} for details). If we denote $\bm{a}^{(l)} = \begin{pmatrix} \phi^{(l)}_*(1) \dotsb \phi^{(l)}_*(N_l) \end{pmatrix}$ as a ``coefficient vector" obtained using the saddle-point, and $\bm{f}^{(l)}$ as the vector with $k$th component representing the relative frequency of the outcome $k$ observed in the experiments for the $l$th measurement setting ($l = 1, \dotsc, L$), then the estimator can be written as
\begin{equation}
    \widehat{F}_*(\bm{f}^{(1)}, \dotsc, \bm{f}^{(L)}) = \sum_{l = 1}^L R_l \ip{\bm{a}^{(l)}, \bm{f}^{(l)}} + c.
\end{equation}

The technical details associated with the method as well as details on how we perform the optimization are given in the accompanying Ref.~\cite{PRA}. Importantly, since this method can be used to estimate any linear functional, we can obtain an estimator for the expectation value of any observable $\mathcal{O}$ by replacing $\rho$ with $\mathcal{O}$ in the above equations. Moreover, any so constructed estimator is experimentally robust in the sense that it does not amplify noise, as we show in the accompanying Ref.~\cite{PRA}.

\paragraph{Examples}
The above definitions of the estimator and risk are very abstract. To get a better feeling for these quantities, we now consider two thought experiments. First, consider a single qubit with target state $\rho = \op{1}{1}$, where $\{\ket{0}, \ket{1}\}$ are the eigenstates of the Pauli-$Z$ operator. We simulate $100$ repetitions of a $Z$ measurement, with outcomes denoted $o_i \in \{0, 1\}$. For a desired confidence level of $95\%$, the above method then gives the following estimator (see section II.B of Ref.~\cite{PRA} for details)

\begin{equation}
    \widehat{F}_*(\{o_i\}_{i = 1}^{100}) = \frac{0.952}{100} \sum_{i = 1}^{100} o_i + 0.024 .
\end{equation}
Hence, $\widehat{F}_*$ turns out to be nothing other than the sample-mean estimator for the Bernoulli parameter, albeit with a small bias to account for finite sample size. Indeed, the problem we have considered is essentially classical, equivalent to estimating the probabilities in a coin flip. As we increase the number of repetitions $R$, we find that the bias goes to zero, while the coefficient multiplying the outcomes approaches $1/R$ as one would expect.

\begin{figure}[ht]
    \includegraphics[width=0.35\textwidth]{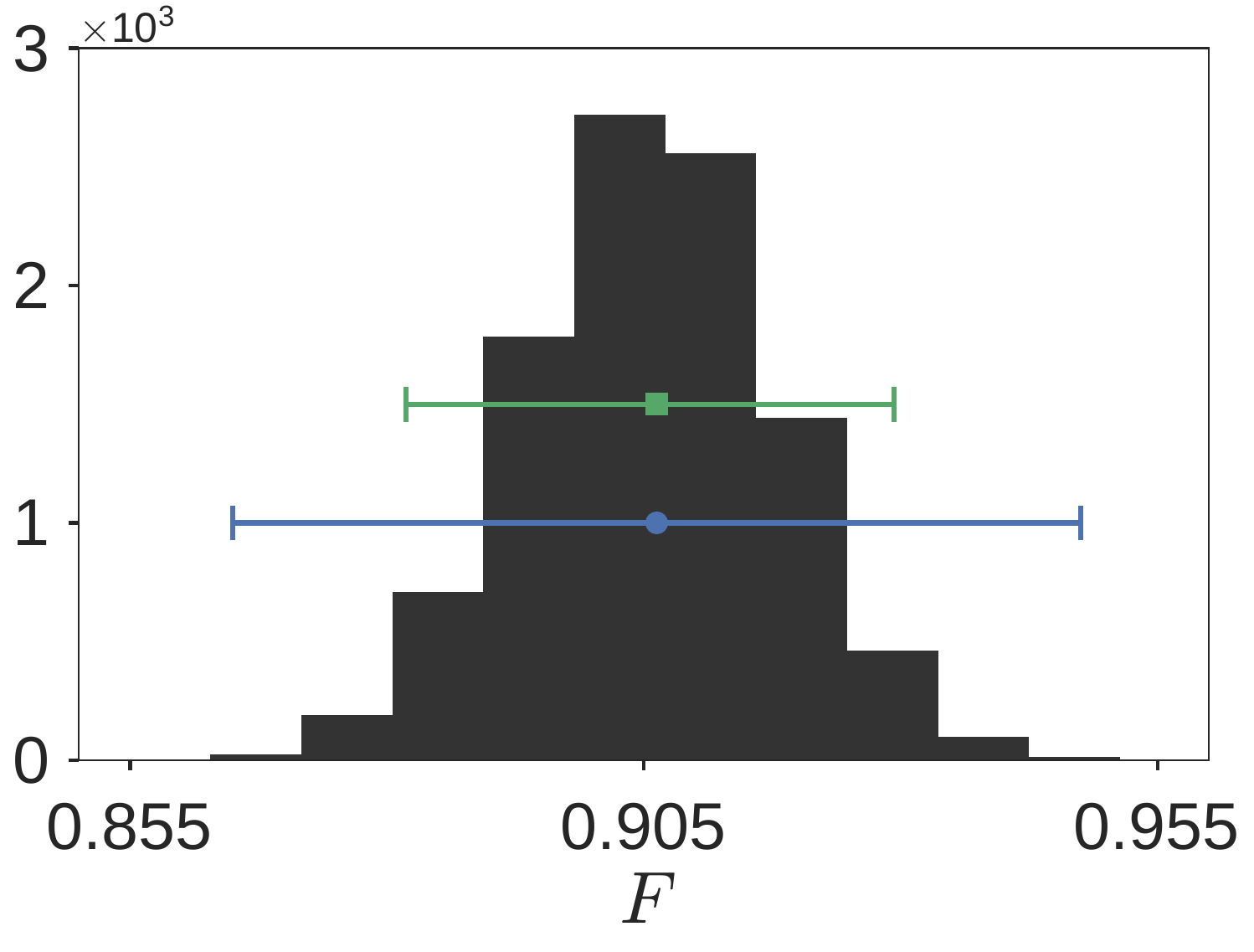}
    \caption{Histogram of $10^4$ estimates obtained using the estimator constructed in Eq.~\eqref{eq:estimator} for a random $4$-qubit target state using outcomes from three-fourths the total number of Pauli measurements. The true fidelity is marked by a green square and a blue dot. The green bar passing through the square indicates the estimated optimal (asymmetric) confidence interval corresponding to a confidence level of $95\%$, while the blue bar passing through the circle indicates the risk associated with the estimator of Eq.~\eqref{eq:estimator}.}
    \label{fig:JN_estimates_hist_risk}
\end{figure}

Next, we consider a random $4$-qubit target state and simulated laboratory state that is obtained by applying $10\%$ depolarizing noise to the target state. We simulate $100$ repetitions of the first $L = 192 = 0.75 \times 4^4$ Pauli measurements (excluding identity) used in DFE~\cite{Flammia2011}, chosen in decreasing order of weights (where 0.75 is an arbitrary cutoff). Figure~\ref{fig:JN_estimates_hist_risk} shows a histogram of the estimates given by the estimator constructed in Eq.~\eqref{eq:estimator}, obtained using $10^4$ sets of simulated outcomes. In order to assess the risk associated with this estimator, we first compute the smallest asymmetric interval around the true fidelity that captures $95\%$ of the estimates~\footnote{This is not technically a confidence interval, since the true fidelity is unknown in practice. The confidence interval would thus be centered around the estimate, and contain the true value with the specified confidence level.}. We find that the confidence interval obtained from the minimax method is only $1.74$ times larger than this optimal asymmetric interval~\footnote{This interval is mostly symmetric for $F \approx 0.9$, but it is expected to be more asymmetric as we approach a fidelity close to $0$ or $1$.}: a small price to pay for a rigorous confidence interval. Note, that the asymmetric interval is always expected to be tighter than the optimal minimax interval because we use the true fidelity to compute it, which would inaccessible in practice. Nevertheless, we find that the above ratio is much better than the guaranteed upper bound of $\vartheta(0.05) = 4.58$ in Eq.~\eqref{eq:theta}.

\paragraph{Experimental data}
We now proceed to testing the estimator of Eq.~\eqref{eq:estimator} on experimental data from a trapped-ion quantum processor~\cite{Schindler2013}. We consider data for three different $4$-qubit target states: a GHZ state, a W-state, and a locally-rotated linear cluster state. Each dataset consists of $81$ Pauli measurements with $100$ shots each (i.e., $R_l=100$ for all $l$). For comparison, we also perform state-of-the-art Maximum Likelihood Estimation (MLE)~\cite{Hradil1997, James2001}. Table~\ref{tab:experimental_data_JN_MLE} shows that both methods produce almost identical estimates, yet the risk for the minimax method is larger than the corresponding risk for the MLE estimate, which is \emph{estimated} using Monte-Carlo (MC) re-sampling of the data. Although smaller, the MC risk for the MLE estimator is not computed from raw data, but requires reconstructing the state which is prone to errors and bias~\cite{Langford2012, Scholten2016, Schwemmer2015QST, Ferrie2018, Blume-Kohout2010BME}. Furthermore, in Ref.~\cite{PRA}, we give an example where the MC-confidence region for MLE is overconfident when provided with incomplete data. In contrast, the more conservative minimax risk, although not as tight, \emph{is guaranteed to be correct}, irrespective of the underlying actual state and the performed measurements.
As this is experimental data, we do not know whether the MC risk estimate is correct.

\begin{table}[!ht]
    \begin{center}
        \begin{tabular}{lcccc}
            \toprule
                & \multicolumn{2}{c}{Minimax method} & \multicolumn{2}{c}{MLE} \\
            \cmidrule(r){2-3} \cmidrule(l){4-5}
                & F Estimate & Risk & F Estimate & MC risk \\
            \midrule
                GHZ & $0.84$ & $0.053$ & $0.84$ & $0.023$ \\
                W & $0.89$ & $0.049$ & $0.88$ & $0.019$ \\
                Cluster & $0.79$ & $0.048$ & $0.79$ & $0.020$ \\
            \bottomrule
        \end{tabular}
    \end{center}
    \caption{Fidelity estimates and risk for a $4$-qubit GHZ state, W state, and a cluster state from experimental data. Estimates are calculated using the minimax method as well as MLE. The risk for MLE is obtained from Monte-Carlo (MC) sampling, which is a heuristic and not guaranteed to be correct. The risk for either method corresponds to a $95\%$ confidence level.}
    \label{tab:experimental_data_JN_MLE}
\end{table}

\paragraph{Stabilizer states}
The experimental data in the previous example used the same set of Pauli measurements for all three target states. However, it is typically desirable to implement measurement schemes that are tailored to a certain target state. A prime example are stabilizer states, which are commonly used in quantum error correction~\cite{Gottesman1997, Kitaev1997}. An $n$-qubit stabilizer state is fully characterized by $n$ stabilizer generators. However, measuring any subset of $n-1$ stabilizer generators is insufficient to determine the state, as that information is still compatible with two orthogonal candidate states (and superpositions thereof). In contrast to MC methods, our method automatically incorporates this fact by returning a risk of $0.5$ when using such a measurement scheme (see Ref.~\cite{PRA}). On the other hand, it is known that measuring the elements of the stabilizer group (excluding the identity) uniformly at random is a minimax optimal measurement strategy (when only Pauli measurements are allowed)~\cite{pallister2018optimal, klieschcharacterization}. We show in Ref.~\cite{PRA} that, with this measurement strategy, our method gives a sample complexity better than or equal to $\approx 2 (\frac{d-1}{d})^2 [\ln(2/\failure)/\jnrisk^2]$, where $d = 2^n$ is the dimension of the system. In other words, for any given stabilizer state, one needs to measure at most a constant number of Pauli operators to estimate the fidelity with a risk of $\jnrisk$ and a confidence level of $1 - \failure$. This sample complexity rivals that of DFE, showing that our method is highly practical when the measurement scheme is suitably chosen for the target state.

\paragraph{Comparison to direct fidelity estimation}
The above measurement scheme can be generalized to arbitrary pure states, resulting in a similar sample complexity and computational speed as DFE. However, under equal conditions with the same measurements and sampling, we find that the our method typically achieves significantly tighter confidence regions than DFE. One reason for this is that DFE uses only the observed eigenvalues ($\pm 1$). In contrast, the minimax method can utilize the additional information about which eigenvectors of the multi-qubit Pauli measurements were observed~\cite{PRA}. This information is typically available in the experiments. 

Moreover, the minimax method is not constrained to using Pauli measurements like DFE specifies. In fact, we show in Ref.~\cite{PRA} that for random target states we can achieve very similar performance regardless of whether we use the DFE measurements or just random POVMs for the same number of shots. This opens the interesting possibility to design measurement schemes that are tailored to a certain target state in the same way that Pauli measurements are tailored to stabilizer states. Depending on the experimental constraints and the complexity of the target state, this can allow for much more efficient estimation. 

We note that our method also has similarities with Quantum State Verification (QSV)~\cite{pallister2018optimal} and Quantum State Certification~\cite{buadescu2019quantum}, which tests whether the fidelity is above a fixed threshold. We show that using the measurement protocol introduced in QSV, we can optimally estimate the fidelity with any $2$-qubit state. We also compare with classical shadows~\cite{Huang2020}, and show that our method can match their sample complexity in principle. While computing our estimator for global Clifford measurements can be computationally challenging, we show that we can get an exponential advantage over classical shadows using appropriate Pauli measurements. Details of these comparisons are included in our companion paper~\cite{PRA}.

\paragraph{Entanglement witnesses}
The minimax method introduced above is not limited to estimating fidelity, but can be applied to arbitrary observables. For example, it is often desirable to estimate the expectation value of an entanglement witness to determine if a (pure or mixed) state is entangled. Consider a Werner state $\rho_{\text{Werner}} = p \frac{\id + \textsc{swap}}{d(d + 1)} + (1{-}p) \frac{\id - \textsc{swap}}{d(d - 1)}$, which is a bipartite state, characterized by a parameter $p \in [0, 1]$ and is entangled for $p < 0.5$~\cite{chruscinski2014entanglement}. The $\textsc{swap}$ operator is an entanglement witness for Werner states~\cite{chruscinski2014entanglement}, but often not easy to measure in practice. For the sake of demonstration, we consider a 2-qubit Werner state with $p = 0.25$ as the target state, and apply $10\%$ depolarizing noise to obtain the actual state, resulting in a true expectation value of $\tr(\textsc{swap}\cdot \sigma) = -0.4$. We perform $3$ Pauli measurements, with $100$ repetitions each, and obtain an estimate of $-0.28$ with a risk of $0.23$. Note that while this risk is quite large, it is guaranteed to be correct. Since $-0.28 + 0.23 < 0$, we can certify with $95\%$ confidence that the state is entangled.

\paragraph{Computational efficiency}
One of the main challenges of the minimax method is computing the estimator efficiently in large dimensions for arbitrary target states and POVMs. Because we need to perform optimization in such situations, quantifying the computational resources required becomes important. We summarize the time and memory consumed to build the estimator for a random target state and a GHZ state in Tab.~\ref{tab:JNprofiling}. Changing the number of repetitions results in negligible overhead. Although the computational time grows exponentially with qubit number (as does the size of the system and number of measurements), structured states give significant improvements. Note that the estimator needs to be constructed only once for a specified measurement scheme and target state. Once the estimator is built, we can give fidelity estimates from raw data almost instantaneously.
This is in contrast to methods such as MLE, which incur high computational cost for every evaluation on the data.

We note that it is possible to simplify the optimization problem depending on the target state and the measurement scheme. For example, we reduce the problem of constructing an estimator for stabilizer states (when measuring random elements of the stabilizer group) to a two-dimensional optimization problem, irrespective of the stabilizer state or the dimension. The resulting algorithm is both time and memory efficient for any number of qubits. The same efficient algorithm can be used for arbitrary target state with a randomized Pauli measurement scheme that we introduce in section II.E of Ref.~\cite{PRA}. However, as in DFE, computing the weights of the different measurements can be cumbersome for complicated states.

\begin{table}[!ht]
    \begin{center}
    \scalebox{0.92}{
        \begin{tabular}{l c c c c c c}
            \toprule
            \makebox[-0.5cm][l]{State}\hspace{1cm} &$n$ & 1 & 2 & 3 & 4 & 5 \\
            \midrule
            \multirow{5}{*}{\rotatebox{90}{Random}} & $L$ & 3 & 12 & 48 & 192 & 768 \\
            \cmidrule{3-7}
            & Nesterov & 49.8 s  & 1.12 min & 5.16 min & 27.7 min     & 2.86 hr$^*$ \\[2pt]
       & \texttt{cvxpy} & 0.088 s & 0.49 s   & 4.53 s   & 1.05 min$^*$ & \rule{0.5cm}{0.4pt} \\[2pt]
   & \texttt{cvxpy-mem} & 0.081 s & 0.69 s   & 8.37 s   & 1.85 min$^*$     & 39.8 min$^{**}$\\[2pt]
            \midrule
            \multirow{5}{*}{\rotatebox{90}{GHZ}} & $L$ & 2 & 4 & 8 & 16 & 32 \\
            \cmidrule{3-7}
            & Nesterov & 2.77 s  & 31.6 s & 3.21 min & 8.75 min & 13.9 min    \\
      & \texttt{cvxpy} & 0.065 s & 0.17 s & 0.68 s   & 4.05 s   & 30.6 s$^*$ \\[2pt]
   & \texttt{cvxpy-mem} & 0.063 s & 0.23 s & 1.27 s   & 8.86 s   & 1.30 min$^*$\\
            \toprule
        \end{tabular}
    }
    \end{center}
    \caption{Time taken to construct the estimator for a random $n$-qubit target state and an $n$-qubit GHZ state with each of our three implementations (average of $3$ simulations). We use $L = 0.75 \times 4^n$ Pauli measurements for the random state, while $L = 2^n$ Pauli measurements for the GHZ state. 
    The computations were performed on a 3.0GHz CPU without parallelization.
    Entries marked with $^*$ require $>$1GB of memory, entries marked with $^{**}$ require $>$32GB, and the \texttt{cvxpy} algorithm run on a random 5-qubit state requires $>$256GB.
    }
    \label{tab:JNprofiling}
\end{table}

\section{Outlook}
We have introduced a method that provides nearly minimax-optimal risk for any specified measurement scheme and target state. The sample complexity of our method is comparable to DFE, when using the same kind of Pauli measurements. However, in contrast to DFE and other methods, we are not imposing any restrictions on the measurement scheme. Hence, depending on the experimental constraints, tailored measurement schemes can be much more efficient than rigid Pauli schemes. In particular, it would be interesting to study the performance of our method for random measurements used in Ref.~\cite{Huang2020}.

The risk is calculated based solely on the measurement scheme, target state, and number of repetitions, so we can use our method as a tool to benchmark experimental protocols. By providing a guarantee on the achievable error for a given protocol, this method can be useful for guiding the experimental design and finding the best protocol for a given target state. Another important avenue for future research is extending the method to estimating nonlinear functions of the state like purity or entanglement measures like negativity. Furthermore, by exploiting the Choi-Jamio\l{}kowski isomorphism, our method can be extended to quantum channels. Since quantum process tomography is even more demanding than state tomography, direct estimation of channel fidelities or other channel properties is extremely useful for characterizing multi-qubit channels.

\section*{Acknowledgements}
The authors thank Emanuel Knill, Scott Glancy, and Yanbao Zhang for helpful discussions on the manuscript.\\
All Python code required to use our method is available at \href{https://github.com/akshayseshadri/minimax-fidelity-estimation}{https://github.com/akshayseshadri/minimax-fidelity-estimation}.\\
This material is based upon work supported by the National Science Foundation under Grant Nos. 1819251 and 2112901.\\
This work utilized the Summit supercomputer, which is supported by the National Science Foundation (awards ACI-1532235 and ACI-1532236), the University of Colorado Boulder, and Colorado State University. The Summit supercomputer is a joint effort of the University of Colorado Boulder and Colorado State University.\\

We gratefully acknowledge support by the Austrian Science Fund (FWF Grant-DOI 10.55776/F71) (SFB BeyondC) and the Institut f\"ur Quanteninformation GmbH. We also acknowledge funding from the EU H2020-FETFLAG-2018-03 under Grant Agreement no. 820495, by the Office of the Director of National Intelligence (ODNI), Intelligence Advanced Research Projects Activity (IARPA), via US Army Research Office (ARO) grant no. W911NF-16-1-0070 and W911NF-21-1-0007, and the US Air Force Office of Scientific Research (AFOSR) via IOE Grant No. FA9550-19-1-7044 LASCEM. This project has received funding from the European Union’s Horizon 2020 research and innovation programme under the Marie Skłodowska-Curie grant agreement No 840450.
\end{document}